\documentclass[twocolumn,showpacs,preprintnumbers,amsmath,amssymb]{revtex4}
\usepackage{graphicx}
\usepackage{dcolumn}
\usepackage{bm}

\begin{document}

\title{\bf{Coarsening Dynamics of Granular Heaplets in Tapped Granular Layers}}

\author{Y.~K.~GOH}
\author{R.~L.~JACOBS}
\affiliation{Department of Mathematics, Imperial College\\
180 Queen's Gate, LONDON SW7 2AZ, U.K.}
\date{\today}


\begin{abstract}
A semi-continuum model is introduced to study 
 the dynamics of the formation of granular heaplets 
 in tapped granular layers.
By taking into account the energy dissipation of collisions  and
 screening effects due to avalanches, 
 this model is able to reproduce qualitatively the pattern of 
 these heaplets.
Our simulations show that the granular heaplets are characterised
 by an effective surface tension which depends on the 
 magnitude of the tapping intensity.
Also, we observe that there is a coarsening effect in that the
 average size of the heaplets, $V$ grows as the number of taps $k$ increases.
The growth law at intermediate times can be fitted by a scaling function
 $V \sim k^z$ but the range of validity of the power law 
 is limited by size effects.
The growth 
 exponent $z$ appears to diverge as the tapping intensity is increased.
\end{abstract}

\maketitle 

In recent years, there has been increasing interest in the collective
 dynamics of granular materials.
The dissipative nature of granular materials gives rise to properties
 distinct from those of solids and liquids.
Many experimental and theoretical
 attempts have been made to seek a fundamental understanding of this 
 granular state and especially pattern formation 
 in driven granular layers.
The experiments include vertically vibrated 
 systems\,\cite{Umbanhowar96,bizon98}, 
 tapped or blown thin films of powders\,\cite{Duran00,Duran01}, 
 and electrostatically driven granular layers\,\cite{Aranson02,Aranson00}.
Although many papers have been published to explain the experiments on
 tapped granular layers,
 most of them deal with either compactification of 
 thick granular layers\,\cite{Ben-Naim98,Knight95}, 
 or static properties of granular heaplets on a thin granular
 layer\,\cite{Duran01}.
The dynamical aspect of the heaping process in a tapped thin granular layer
 is still not well understood.
This paper is concerned with the coarsening dynamics of 
 granular heaplets in tapped granular layers from 
 a theoretical point of view.

This paper is organised as follows.
First we introduce a simple model to study this fascinating system
 by considering the energy dissipation  and screening effects during
 the tapping process.
Despite the simplicity, this model is capable 
 of capturing the essential phenomenology, 
 reproducing various morphologies of the coarsening pattern, 
 and showing the way in which heaplets merge 
 as observed in experiments\,\cite{Duran01}. 
The model is then studied numerically and the results indicate some of
 the relationships between tapping intensity and
 the effective surface tension of granular layers.
Finally analysis of 
 the results shows that the average size of the the heaplets $V$ grows 
 as a power law with the number of taps $k$, $V\sim k^z$
 for limited range of $k$ values.
The exponent $z$ appears to diverge as the tapping intensity is increased.

{\em Model} -- 
Consider a {\em thin} layer of $N \gg 1$ granular particles
 spread out over a flat plate.
The plate is then tapped repeatedly at a low pace
 with constant shock intensity.
In one complete tapping cycle, there are two different processes 
 each of which requires a different description.
In the tapping phase,
 the granular layer is perturbed by an external shock and
 granular particles acquire kinetic energy to allow 
 them explore the phase space and hop on to neighbouring sites.
The hopping range of the granular particles depends on
 the amount of kinetic energy received by each of 
 the individual particles.
The amount of energy supplied to the system is controlled by
 the dimensionless acceleration $\Gamma=U/g \, \Delta t$,
 where $U$ is the velocity of the plate when it is in motion and
 $\Delta t$ is the time interval over which it is in motion.
Note that $U$ is also the vertical velocity gained by 
 an {\em elastic} particle sitting on the plate when it is tapped.
Our particles are not however elastic.
Therefore, the vertical velocity gain is $\alpha U$,
 where $0 \le \alpha \le 1$ is the departure coefficient
 analogous to the coefficient of restitution and 
 it characterises the degree of dissipation of the system. 
After the tap, the particles undergo a ballistic flight and fall back 
 again onto the static plate where 
 they relax subject to avalanche dynamics before the next tap.
In this relaxation phase, the granular particles may stay immobile
 or move about depending on the local slope.
If the local slope is less than a critical slope then the profile 
 remains stationary.
Conversely,
 if the local slope is greater than the critical slope,
 matter moves down the slope collectively as an avalanche, 
 until the slope is less than the critical slope.

Let the area density $n(\mathbf{x},k)$ be the number of
 granular particles above a unit area of the plate after $k$ taps.
Here $\mathbf{x} = (x,y)$, where $x$ and $y$ are 
 orthogonal coordinates parallel to the plate.
The average diameter, $D$ of the granular particles is chosen
 to be $1$, so that $n( \mathbf{x},k)$ now is equal to the 
 local thickness of the granular layer, so long as there is no
 compactification throughout the tapping process.
Of course this is just an idealisation, 
 the packing fraction of the granular material will be 
 changed\,\cite{Nowak98} if the granular layer is thick 
 or it is intentionally prepared in a low-density state, but 
 such cases will not be considered here.

We can approximate the acceleration $A_p$ of the plate's movement
 due to tapping as a sum of $M$ delta functions
 \begin{equation}\label{eq:A_p}
   A_p(t) = \sum_{k=1}^{M}U\, \delta(t-(k-1)\tau),
 \end{equation}
 where the velocity $U$
 gives the intensity of the shocks which are assumed uniform
 over the whole plate. $\tau$ specifies the time interval between 
 taps. 
Here $\tau$ must be greater than any time scale in the relaxation
 process, this is to ensure that next tap only occurs
 after the system is fully relaxed.
In the dilute limit, a single inelastic grain sitting on the plate
 experiences a net force 
 \begin{equation}\label{eq:F_z}
  F_z(t) = \alpha \left( \sum_{k=1}^{M}mU\, \delta(t-(k-1)\tau)\right) - mg,
 \end{equation}
 where $\alpha$ is the departure coefficient mentioned earlier.
During a small interval 
 $[t_k - \Delta t/2, t_k + \Delta t/2]$, 
 where $t_k$ is the short hand for $t_k = (k-1)\tau$, the time
 of $k$-th tap, and the maximum velocity transferred to the particle is
 \begin{equation}\label{eq:v_z}
   \Delta v_z = \int_{t_k - \frac{\Delta t}{2}}^{t_k + \frac{\Delta t}{2}}
       \frac{F_z(t)}{m} \,  \mathrm{d}t = 
   \alpha U - g \, \Delta t.
 \end{equation}
In order for the particle to hop, $\Delta v_z$ must be positive,
 or $\alpha U > v_0 \equiv g\, \Delta t$
 so that external acceleration overcomes the gravitational force.
In what follows this is always the case.

In the continuum limit one has to make two crucial modifications.
$\alpha$ must be replaced by an {\em effective} departure coefficient
 of the granular bulk. 
$\alpha$ is expected to be a monotonic decreasing function of 
 $n(\mathbf{x},k)$.
This is because when the number density is high,
 inter-grain collisions occur more often and more energy is dissipated,
 so that grains depart with a smaller departure velocity.
While the precise form of $\alpha(n)$ is unknown, in this 
 paper it is taken to be 
 \begin{equation}\label{eq:alpha}
   \alpha(n) = \left\{
    \begin{array}{ll}
       \frac{\alpha_0 \,\bar{n}}{n(\mathbf{x},k)},
     & n > 1 
     \\
       \alpha_0 \bar{n},
     & n \le 1,
     \\ 
    \end{array}
    \right.
 \end{equation}
 for reasons of simplicity and to make comparison with Duran's 
 model\,\cite{Duran01}.
Here, $\bar{n}$ is the average density of the system and
 $\alpha_0$ is the effective departure coefficient for
 average density $\bar{n}$. 
Also, the second term in Eq.(\ref{eq:F_z}) needs to be changed.
In the continuum limit, according to Duran\,\cite{Duran00}
 there is an effect dependent on the position of a grain in the heap.
If the grain is not at the top of the heap it supports a fraction of the weight
 of the grains above it, 
 hence its effective mass is increased.
As a result, in order for a granular particle sitting on the inclined 
 side of the granular layer to hop,
 it requires a larger velocity kick than a particle sitting
 on the flat region of the layer.
This screening effect can be incorporated into Eq.(\ref{eq:F_z}) by
 replacing $m$ in the second term with an effective mass
 \begin{equation}\label{eq:mass}
   m^* = \left((n_T - n) p \,\sin \theta_c /D \right) m.
 \end{equation}
Here $D$ is the average grain diameter and it is set to 1 henceforth.
$n_T$ is the altitude of the nearest peak in the corrugated granular layer.
$p$ is a parameter of unknown value which is set equal to 5 in 
 reference~\cite{Duran00}
 and this value appears to give a match to experimental results.
$\theta_c$ is the angle of repose 
 and is close to the value of $\pi/6$ \,\cite{Duran00}.
After these modifications, Eq.(\ref{eq:v_z}) now can be written as
 \begin{equation}\label{eq:dv}
 \frac{\Delta v_z}{v_0} = \left\{
  \begin{array}{ll}
     \frac{\mathcal{A}}{n} - (n_T - n)p\,\sin\theta_c
   & n > 1 
   \\
     \mathcal{A}- (n_T - n)p\,\sin\theta_c
   & n \le 1,
   \\
  \end{array}
  \right.
 \end{equation}
where $\mathcal{A} = \frac{\alpha_0 \bar{n} U}{v_0}$ is 
 the tapping intensity.

There are close similarities between our model and Duran's~\cite{Duran01}.
In both models, the resulting pattern formation is due to the competition 
 between the upward hopping motion of particles and the downward screening 
 effect of avalanches.
However, the mechanism causing upward motion is different in the two models. 
Duran~\cite{Duran01} and Shinbrot~\cite{Shinbrot98} 
 conjectured that the hopping of granular particles is due to the
 upcoming air-flux through the porous bed of the granular layer.
The velocity of the air-flux at height $n$ can be approximated by
 Darcy's law $v_{air}=Kp/n$, 
 where $K$ is the permeability of the granular bed, 
 $p$ is the pressure different across the granular bed.
The velocity of the air-flux $v_{air}$
 must be greater than the free fall velocity
 of a granular particle $v_{f}$ in order to eject this particle.
However, in our model we adopt a simpler picture where
 kinetic energy is transferred to the granular  particles 
 via direct collisions between plate and particle and
 between particle and particle.
Energy is dissipated via the effective departure coefficient.
The form of the effective departure coefficient is chosen as
 in Eq.(\ref{eq:alpha}) so that the velocity transfer from plate to 
 the particle
 has the same $1/n$ dependency ($n \ge 1$) as $v_{air}$ in Darcy's law.
Obviously, we have the freedom to choose other functional forms of $\alpha(n)$,
 the simple choice here is just to achieve comparison with Duran's model.
As observed from simulation, however, it turned out that the exact form of 
 $\alpha(n)$ is unimportant so long as it is a monotonic decreasing function.

Now it remains to specify the hopping process in the tapping phase.
When the granular layer is tapped, at an unstable site where $\Delta v_z > 0$,
 particles can hop to 
 any neighbouring site within a circle defined by a maximum
 horizontal hopping range, $R_{max}({\mathbf \Delta v})$.
Here we assume on the unstable site,
 all the granular particles are re-distributed.
Each grain hops with equal probabilities in any direction and
 with a random hopping range, $R<R_{max}$.
We can estimate the maximum hopping range
 $R_{max}({\mathbf \Delta v})$ by the following 
 simple arguments.
The maximum velocity received by a particle during one tap is 
 $\Delta v_z$ vertical to the plate.
In this case $R_{max}({\mathbf \Delta v})$ is zero
 for the motion is strictly vertical.
It is unlikely that particles will always hop in the 
 vertical direction, an inter-particle collision can change the
 hopping direction to any angle.
Assume that particle is ejected with a 
 speed $\Delta v_z \cos\phi$ with $\phi$ denoting 
 the angle from the vertical axis of the plate.
Then, the horizontal hopping range is
 \begin{equation}\label{eq:R}
   R = \Delta v_z \cos\phi \sin\phi \,t'
     = \frac{2 \cos^3\!\phi \sin\phi}{g} (\Delta v_z)^2,
 \end{equation}
where $t'$ is the flight time of the particle in vacuum.
$R$ is maximum when $\phi = \pi/6$, so that the estimated value for $R_{max}$
 is given by
 \begin{equation}\label{eq:R_max}
   R_{max}(\Delta v_z) = \frac{3\sqrt{3}}{8g}(\Delta v_z)^2
                       = \xi (\frac{\Delta v_z}{v_0})^2
 \end{equation}
 where 
 \begin{equation}\label{eq:xi}
   \xi = \frac{3\sqrt{3}}{8}v_0\,\Delta t.
 \end{equation}
The estimated $R_{max}$ is very crude and does not take into
 account the details of the air-particle and particle-particle interactions
 except phenomenologically.
However, one can consider $\xi$ in $R_{max}$ as an adjustable
 parameter, which varies from one experiment to another experiment
 depending on the roughness of the material used in experiment
 and humidity of the surrounding environment.

There are several ways to describe the avalanche process 
 during the relaxation phase.
Here we use a slope dependent diffusion equation.
We start from the equation of continuity
 $\partial_t{n} = - \mathbf{ \nabla \cdot J}$ with the constitutive
 current $\mathbf{J}$ given by
 \begin{equation}\label{eq:J}
   \mathbf{J} = \left\{
   \begin{array}{ll}
     -\eta(\beta |\nabla n|^2 - 1)\nabla n, 
     &
     |\nabla n| > \frac{1}{\sqrt{\beta}} 
     \\
     0, &
     |\nabla n| \leq \frac{1}{\sqrt{\beta}} ,
   \end{array}
   \right.
 \end{equation}
 where $1/\sqrt{\beta} = \tan\theta_c$ is the critical slope
 and $\eta$ sets the diffusion rate.
The current is chosen in such a way that when the gradient is 
 greater than the critical slope, mass current flows down-hill
 to smooth out density fluctuations, 
 but no mass is transferred if the gradient is less than the critical slope.

{\em Simulation} -- 
The model is studied on a $N \times N$ square lattice with a periodic
 boundary condition.
Initially, the granular layer is prepared by assigning a random height
 between $[0,5]$ to each site and letting the granular layer relax.
At the beginning, the fluctuation in $n(\mathbf{x},0)$ is considerably 
 smaller then the height of the heaplets formed later.
On each tapping cycle, the velocity of particles at each site is calculated 
 and the density number is updated according to 
 rules defined by Eq.(\ref{eq:dv}) and Eq.(\ref{eq:R_max}).
After which the equation of continuity and 
 the corresponding constitutive current 
 equation Eq.(\ref{eq:J}) are solved and iterated numerically until 
 there is no more mass is transferred at each site.

\begin{figure}[!htbp]
 \vspace{0.3cm}
 \includegraphics[width=8cm]{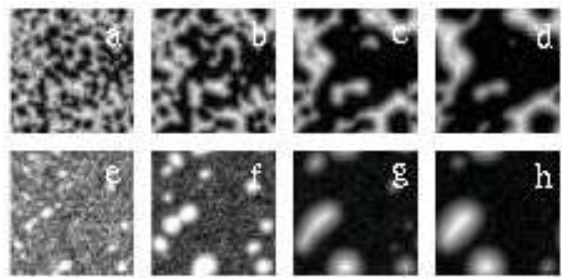}
 \caption{\label{fig:densityPlot}
  Two simulations of coarsening dynamics of tapped granular layers
  corresponding to different tapping intensities.
  The simulation images (a)--(d) and (e)--(h) start with
  the same initial configuration.
  Figures (a)--(d) use a smaller tapping intensity
    ($\mathcal{A}=6.0, \xi=0.2207$), 
    and Figures (e)--(h) correspond to a stronger tapping intensity 
    ($\mathcal{A}=16.8, \xi=0.2207$).
  (a) and (e) are taken at $k=5$, 
  (b) and (f) at $k=10$,
  (c) and (g) at $k=20$,
  (d) and (h) at $k=30$.
 }
\end{figure}

The following density plots show typical results from the simulation.
Fig.\ref{fig:densityPlot} shows the density plots 
 of $n(\mathbf{x},k)$ at different values of $k$, the number of taps,
 for two set of parameters.
The top panel shows extended patterns of ridges and
 corresponds to a smaller tapping intensity ($\mathcal{A}=6.0$), 
 while the lower panel shows localised heaplets and corresponds to 
 a greater tapping intensity ($\mathcal{A}=16.8$).
Both simulations coarsen when $k$ increases, eventually 
 reaching their final stage where the entire system is a single 
 granular heap.

One can see from Fig.\ref{fig:densityPlot}
 that as the tapping intensity is increased, 
 the patterns favour isolated circular heaplets 
 which suggests that an effective surface tension can be defined
 which increases with the tapping intensity.
Since there are no attractive forces between granular particles,
 the granular layer cannot have a true surface tension.
This ``surface tension'' effect is entirely due to the system trying to
 maximise the local energy dissipation by decreasing 
 the value of the effective departure coefficient 
 so that particles are less mobile.
In order to maximise the local energy losses, 
 particles have to be as close to each other as possible, 
 for this will increase the
 number of inelastic collisions between particles.
Therefore, with the constraint that the local slope should
 not exceeds the critical slope,
 the global attractor of the pattern is a single large heap, 
 where local energy loss is the largest.
Akiyama {\em et al.}\,\cite{Akiyama98},
 Cl\'ement\,\cite{clement92} and Duran\,\cite{Duran01}
 have discussed the effects of convection on heaping process of 
 granular layers.
Duran\,\cite{Duran01} in particular suggests that 
 the ``surface tension'' is brought about by the convective forces
 dragging particles from the surroundings into the granular heaplets. 
However, in our simulation it is due to the maximisation  of energy losses, 
 since in our model the granular layer does not contain any information
 about the internal interactions of the sand heap,
 therefore convective forces are not taken into account.

\begin{figure}[!htbp]
 \vspace{0.3cm}
 \includegraphics[width=8cm]{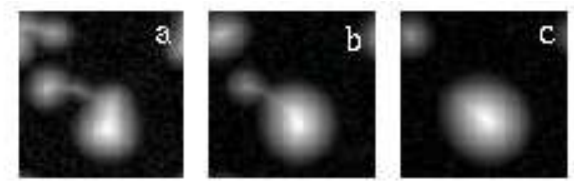}
 \caption{\label{fig:merge}
   Merging of two granular heaplets. $\mathcal{A}=16.0, \xi=0.2207$.
   When two heaplets touch each other, smaller heaplet is sucked into 
   the larger one.
   (a) $k=15$ two granular heaplets meet.
   (b) $k=30$ smaller heaplet is sucked into the larger one.
   (c) $k=50$ small heaplet disappears.
 }
\end{figure}

Nevertheless, the simulation result does 
 show the Laplace-Young pressure effect suggested by reference\,\cite{Duran01}.
A typical example is shown in Fig.\ref{fig:merge}.
(The parameters here are $\mathcal{A}=16.0$ and $\xi=0.2207$,
 but similar results are found over a wide range of 
 values of these parameters.)
In the center of the figure there are two connected granular heaplets.
After several taps the smaller heaplet merges with
 the larger heaplet due to matter moving along the connecting neck.

We are also interested in the dynamics of the size of the heaplets.
The relevant length scale measure $l$ is the average 
 thickness of the site $h_i$ weighted by the local volume, $V_i$,
 and the average volume of the heaplet is $V=l^3$.
Since we know that $h_i=n_i$ and the local volume of the site is just 
 $V_i = n_i \sigma$, where $\sigma$ is the area element of the lattice site,
 it follows that
 \begin{equation}\label{eq:length}
   l = \frac{\sum_i \,h_i\, V_i}{\sum_i V_i} 
     = \frac{\sum_i n_i^2}{\sum_i n_i}.
 \end{equation}

\begin{figure}[!htbp]
 \vspace{0.6cm}
 \includegraphics[width=8cm]{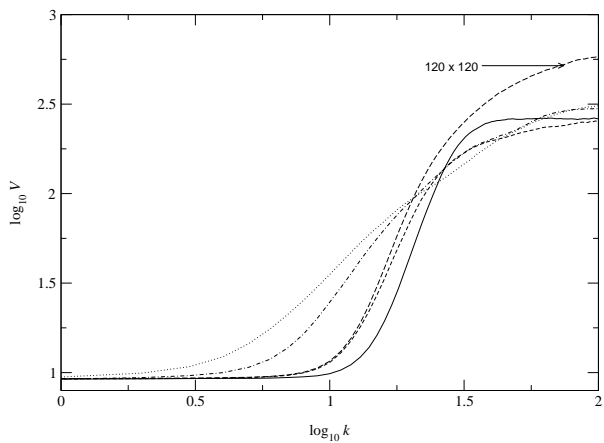}
 \caption{\label{fig:LogLog}
  Log-log plot for characteristic heaplet size $V$ against number of taps $k$
    for varying tapping intensities $\mathcal{A}$.
  The solid line corresponds to $\mathcal{A} = 18.0$,
    the two dashed lines to $\mathcal{A} = 17.6$ 
    the dash-dotted line to $\mathcal{A} = 16.0$,
    and the dotted line to $\mathcal{A} = 14.4$.
  The parameter used in the simulation is $\xi = 0.2207$.
  All curves correspond to a lattice size $L_1 \times L_1 = 80 \times 80$
    and are averaged over ten different runs except for
    the marked dashed curve which corresponds to 
    $L_2 \times L_2  = 120 \times 120$ and is averaged over five runs.
  Each curve saturated after a long time, 
    this can be clearly seen from the $\mathcal{A}=18.0$ solid line.
  The marked curve corresponds to the larger lattice size 
   shows a wider range of the 
   linear region and a larger saturation value of $V$.
 }
\end{figure}

There is some limited evidence for power law behaviour $V \sim k^z$
  in the intermediate region of the log-log plots of $V$ versus $k$ in
  Fig.\ref{fig:LogLog}.
There are three regions in each plot.
The early region is where the granular layer is randomly distributed, 
 the number density fluctuation is not large enough to trigger coalescence.
The late-time region is where all the heaplets have joined into one big heap,
 leaving some remaining individual particles
 hopping randomly about and occasionally encountering the large
 single heap.
The intermediate region is also the fast-growing region where heaplets
 grow and merge into each other.
The range of the intermediate region grows and remains linear as the system 
 size increases as can be seen from the two curves with 
 $\mathcal{A}=17.6$.
This shows that system size limits the range of the intermediate 
 power law region.
For very large system sizes we would expect on this basis that
 the intermediate power law region would have a greater 
 range and remain linear in the log-log plot.
One can easily estimate the increase in
 the average volume of the heaplets $V$ by the 
 following simple argument.
As the saturated value of $V$ corresponds to a state where a single heap 
 contains approximately all the grains in the system, 
 it should be proportional to $\bar{n} L^2$,
 here the average number of  grains per site $\bar{n}$ is a conserved quantity.
Then, increment in $V$ is given by $\Delta\log_{10}V =2\,\Delta\log_{10} L$.
In Fig.\ref{fig:LogLog} 
 the dashed line corresponds to
 a lattice size $L_1 \times L_1=80 \times 80$ and 
 the marked-dashed line to $L_2 \times L_2 = 120 \times 120$,  
 and the vertical shift in the graphs is expected to be
 $\Delta \log_{10}V = 0.352$
 which may be compared to $0.360$ from the simulation.

\begin{figure}[!hbp]
 \vspace{0.5cm}
 \includegraphics[width=7cm]{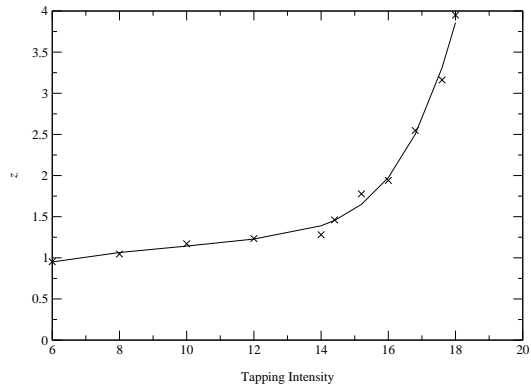}
 \caption{\label{fig:slope_A}
  Exponent $z$ for varying tapping intensity $\mathcal{A}$.
  For small $\mathcal{A}$ ($\mathcal{A}< 14$), $z$ is roughly constant 
  ($z \sim 1.0$), 
  but appears to diverge for $\mathcal{A} > 14$.
  Each value of $z$ is averaged over ten different runs.
 }
\end{figure}
The exponent $z$ in the power law depends on the tapping 
 intensity $\mathcal{A}$ as is shown in Fig.\ref{fig:slope_A}
As we can see $z$ is roughly constant, $z\sim 1.0$,
  for low tapping intensity. 
When the tapping intensity is increased $\mathcal{A}>14$,
 $z$ suddenly increases,  and appears to diverge
 near $\mathcal{A}=18$.
For these large values of the tapping intensity, we observed that the system
 appears to be in a flat configuration for some time 
 and suddenly a single heap formed within a few taps.
For still greater tapping intensities,
 the heaplet patterns disappear and the layer remains flat
 with a small random variation superimposed.
This pattern-disorder transition occurs at a not very precisely defined
 critical tapping intensity of $\mathcal{A} \sim 18$. 
(This critical intensity appears to depend slightly on the 
 initial conditions which is the reason for our statement that 
 it is not precisely defined.)
Above this critical tapping intensity, no patterns are observed.
This is due to the fact that the perturbation is so strong that 
 no site is stable, 
 and each site constantly undergoes re-distribution on each tapping cycle.
Although the power law parametrisation is based on limited 
 evidence, it does provide a useful parameter $z$ to distinguish the 
 pattern-forming region ($\mathcal{A} < 18$) from the non-pattern-forming
 region ($\mathcal{A}>18$).

We have introduced a simple model of the heaping process 
in a tapped granular layer.
The model is capable of reproducing the essential morphologies of tapped
 granular systems.
Qualitatively the effective surface tension of the granular heaps 
 is closely related to the tapping intensity,
 and it is shown that there is no need for convective forces 
 for this effective surface tension to exist.
The length scale of the system coarsens according to 
 the power law $l\sim k^z$ in a limited range.
The exponent $z$ appears to diverge as tapping intensity is increased
 and provides a useful parameter for distinguishing the pattern-forming 
 and the non-pattern-forming regions.

We thank Philip Cheung for helpful discussions.

\end{document}